\documentclass[twoside,a4paper]{article}
\usepackage{amsmath,graphicx,amssymb,fancyhdr,amsthm,enumerate,textcomp} 
\usepackage[usenames]{color}
\pdfoutput=1
\newtheorem{thm}{Theorem}[section]

\newtheorem{prop}[thm]{Proposition} 
 
\theoremstyle{definition} 
\newtheorem{defn}[thm]{Definition}
  
\theoremstyle{remark}  
\newtheorem{rem}[thm]{Remark}  
\def\beq{\begin{eqnarray}}  
\def\eeq{\end{eqnarray}}  
\def\bsp{\begin{split}}  
\def\esp{\end{split}}

\def\d{\mathrm{d}}

\newcommand{\mbold}[1]{\mbox{\boldmath{\ensuremath{#1}}}}

\begin{document}  
  
\title{\Large\textbf{On the algebraic classification of pseudo-Riemannian spaces }}  
\author{{\large\textbf{Sigbj\o rn Hervik$^\text{\tiny\textleaf}$ and Alan Coley$^\heartsuit$} }    
%EndAName    
%\address{    
 \vspace{0.3cm} \\    
$^\text{\tiny\textleaf}$Faculty of Science and Technology,\\    
 University of Stavanger,\\  N-4036 Stavanger, Norway     
\vspace{0.2cm} \\ 
$^\heartsuit$Department of Mathematics and Statistics,\\ 
Dalhousie University, \\ 
Halifax, Nova Scotia, Canada B3H 3J5 
%\email{    
\vspace{0.3cm} \\     
\texttt{sigbjorn.hervik@uis.no, aac@mathstat.dal.ca} }    
\date{\today}    
\maketitle  
\pagestyle{fancy}  
\fancyhead{} % clear all header fields  
\fancyhead[EC]{S. Hervik \& A. Coley}  
\fancyhead[EL,OR]{\thepage}  
\fancyhead[OC]{Algebraic classification of pseudo-Riemannian metrics}  
\fancyfoot{} % clear all footer fields  
  
\begin{abstract} 

We consider arbitrary-dimensional pseudo-Riemannian
spaces of signature $(k,k+m)$.
We introduce a  boost-weight decomposition and define
a number of algebraic properties (e.g., the 
${\bf S}_i$- and ${\bf N}$-properties)
and present a boost-weight decomposition to classifiy the Weyl tensors 
of  arbitrary signature and discuss degenerate algebraic types
(e.g., VSI spaces). We consider the
four dimensional neutral signature space as an illustration.
 
\end{abstract} 

\section{Introduction}

In Lorentzian geometry it is useful to use the boost-weights to
categorise tensors \cite{class,degen,VSI}.  This was particularly useful in studying
degenerate metrics and it is thus useful to generalise this
to the pseudo-Riemannian case \cite{OP,Law}.  
In this short paper we will consider an arbitrary-dimensional pseudo-Riemannian
space of signature $(k,k+m)$.
The symmetry group of
frame-rotations in this case is  $SO(k,k+m)$.  We will utilise the
decomposition where any element, $G$, can be written $G=KAN$, where
we have split it into an compact spin piece, $K$, an Abelian boost
piece, $A$, and a piece consisting of null-rotations $N$.  For
$SO(k,k+m)$, $K\in SO(m)$, and there are $k$-independent boosts (which equals
the real rank of $SO(k,k+m)$).  We want to utilise this
decomposition and we can do so by choosing a suitable frame.

Therefore, we introduce a null-frame such that the metric can be written:
\beq
\d s^2=2\left({\mbold\ell}^1{\mbold n}^1+\dots+{\mbold\ell}^I{\mbold n}^I+\dots+{\mbold\ell}^{k}{\mbold n}^{k}\right)+\delta_{ij}{\mbold m}^i{\mbold m}^j,
\eeq
where the indices $i=1,\dots, m$. The spins will act on ${\mbold m}^i$,  each boost will 
act on the pair of null-vectors, while the null-rotations will in general mix up null-vectors and spatial vectors. More precisely, 

\beq
\text{Spins:} && \tilde{\mbold\ell}^I=\ell^I, ~\tilde{\mbold n}^I={\mbold n}^I, ~\tilde{\mbold m}^i=M^i_{~j}{\mbold m}^j, ~ (M^i_{~j})\in SO(m), \\
\text{Boosts:} && \tilde{\mbold\ell}^I=e^{\lambda_I}\ell^I, ~\tilde{\mbold n}^I=e^{-\lambda_I}{\mbold n}^I, ~\tilde{\mbold m}^i={\mbold m}^j, 
\eeq
while the null-rotations can be split up at each level. Considering 
the subset of forms $({\mbold\ell}^I,{\mbold n}^I,{\mbold\omega}^{\mu_I})$, 
where ${\mbold\omega}^{\mu_I}=\{{\mbold\ell}^{I+1},{\mbold n}^{I+1},\cdots,{\mbold\ell}^{k},{\mbold n}^{k},{\mbold m}^{i}\}$, 
we can then consider the $I$-th level null-rotations w.r.t. ${\mbold n}^I$:
\beq
\text{Null-rot:} ~~ \tilde{\mbold\ell}^I={\mbold\ell}^I-z_{\mu_I}{\mbold\omega}^{\mu_I}-\frac 12z_{\mu_I}z^{\mu_I}{\mbold n}^I, ~\tilde{\mbold n}^I={\mbold n}^I, ~\tilde{\mbold\omega}^{\mu_I}={\mbold\omega}^{\mu_I}+z^{\mu_I}{\mbold n}^I, 
\eeq
and similarly for ${\mbold \ell}^I$. Note that there are 
$2(2k+m-2I)$ null-rotations at $I$th level, making $2k(k+m-1)$ in total.  

\section{Boost-weight decomposition and the ${\bf S}_i$- and ${\bf N}$-properties}

First we need to introduce some mathematical tools which are useful for studying 
these metrics. 
Consider the $k$ independent boosts:
\beq
({\mbold \ell}^1,{\mbold n}^1)&\mapsto& (e^{\lambda_1}{\mbold\ell}^1,e^{-\lambda_1}{\mbold n}^1)\nonumber\\
({\mbold \ell}^2,{\mbold n}^2)&\mapsto& (e^{\lambda_2}{\mbold\ell}^2,e^{-\lambda_2}{\mbold n}^2)\nonumber\\
& \vdots & \nonumber\\
({\mbold\ell}^{k},{\mbold n}^{k})&\mapsto& (e^{\lambda_k}{\mbold\ell}^{k},e^{-\lambda_k}{\mbold n}^{k}).
\eeq
For a tensor $T$, we can further consider the boost weight of the components of this tensor as ${\bf b}\in \mathbb{Z}^k$, as follows. If the component $T_{\mu_1...\mu_n}$ transforms as:
\[
T_{\mu_1...\mu_n}\mapsto e^{-(b_1\lambda_1+b_2\lambda_2+...+b_k\lambda_k)}T_{\mu_1...\mu_n},
\]
then we will say the component $T_{\mu_1...\mu_n}$ is of boost weight ${\bf b}\equiv (b_1,b_2,...,b_k)$. We can now decompose a tensor into boost weights, in particular, 
\[ T=\sum_{{\bf b}\in  \mathbb{Z}^k}(T)_{\bf b},\] 
where $(T)_{\bf b}$ means the projection onto the components of boost weight ${\bf b}$. 

By considering tensor products, the boost weights will obey an additive rule: 
\beq
(T \otimes S)_{{\bf b}}=\sum_{\tilde{\bf b}+\hat{\bf b}={\bf b}}(T)_{\tilde{\bf b}}\otimes (S)_{\hat{\bf b}}.
\eeq

Let us consider a tensor, $T$, and list a few conditions that the tensor components may fulfill:
\begin{defn} \label{cond}We define the following conditions:
\begin{enumerate}
\item[B1)]{} $(T)_{\bf b}=0$, for ${\bf b}=(b_1,b_2,b_3,...,b_k)$, $b_1>0$. 
\item[B2)]{} $(T)_{\bf b}=0$, for ${\bf b}=(0,b_2,b_3,...,b_k)$, $b_2>0$. 
\item[B3)]{} $(T)_{\bf b}=0$, for ${\bf b}=(0,0,b_3,...,b_k)$, $b_3>0$.
\item[$\vdots$]{} 
\item[B$k$)]{}  $(T)_{\bf b}=0$, for ${\bf b}=(0,0,...,0,b_k)$, $b_k>0$.
\end{enumerate}
\end{defn}

\begin{defn}
We will say that a tensor $T$ possesses the ${\bf S}_1$-property 
if and only if there exists a null frame such that condition (B1) above is satisfied. 
Furthermore, we say that $T$ possesses the ${\bf S}_i$-property iff 
there exists a null frame such that conditions B1)-B$i$) above are satisfied.
\footnote{Clearly, we assume that any trivial reordering (i.e., 
relabelling the $b_i$) has been affected.}
\end{defn}
\begin{defn}
We will say that a tensor $T$ possesses the ${\bf N}$-property if and only if 
there exists a null frame such that conditions B1)-B$k$) in definition \ref{cond} are satisfied, \emph{and} 
\[ (T)_{\bf b}=0, \text{ for }  {\bf b}=(0,0,...,0,0).\] 

\end{defn}

\begin{prop}
For tensor products we have:
\begin{enumerate}
\item{}
Let $T$ and $S$ possess the ${\bf S}_i$- and ${\bf S}_j$-property, respectively. 
Assuming, with no loss of generality, that $i\leq j$, then $T\otimes S$, and any contraction thereof, possesses the ${\bf S}_{i}$-property.
\item{} Let  $T$ and $S$ possess the ${\bf S}_i$- and ${\bf N}$-property, respectively. Then  $T\otimes S$, and any contraction thereof, possesses the ${\bf S}_i$-property. If $i=k$, then   $T\otimes S$ possesses the ${\bf N}$-property.
\item{}  Let  $T$ and $S$ both possess the ${\bf N}$-property. Then  $T\otimes S$, and any contraction thereof, possesses the ${\bf N}$-property.
\end{enumerate}
\end{prop}

It is also useful to define a set of related conditions. Consider 
a tensor $T$ that does not necessarily meet any of the conditions above. However, since the boost weights ${\bf b}\in \mathbb{Z}^k\subset{\mathbb R}^k$, we can consider a linear $GL(k)$ transformation, $G:\mathbb{Z}^k\mapsto \Gamma$, where $\Gamma$ is a lattice in $\mathbb{R}^k$. Now, if there exist a $G$ such that the transformed boost weights, $G{\bf b}$, satisfy (some) of the conditions in Def.\ref{cond}, we will say, correspondingly, that $T$ possesses the ${\bf S}^G_i$-property. Similarly, for the ${\bf N}^G$-property.

If we have two tensors $T$ and $S$ both possessing the ${\bf S}_i^G$-property, with the same $G$, then when we take the tensor product: 
\[ (T\otimes S)_{G{\bf b}}=\sum_{G\hat{\bf b}+G\tilde{\bf b}=G{\bf b}}(T)_{G\hat{\bf b}}\otimes(S)_{G\tilde{\bf b}}.\]
Therefore, the tensor product will also possess the  ${\bf S}_i^G$-property, with the same $G$. This would be useful for us later when considering degenerate tensors and metrics with degenerate curvature tensors. Note also that ${\bf S}_i^G$-property reduces to the ${\bf S}_i$-property for $G=I$ (the identity).

\begin{rem}
A tensor $T$ satisfying the ${\bf S}_i^G$-property or ${\bf N}^G$-property 
is generically not determined by its invariants in the sense that there may
 be another tensor $T'$ with the same invariants. The ${\bf S}_i$-property 
 therefore implies a certain \emph{degeneracy} in the algebraic structure of the tensor. 
\end{rem}

\section{Boost-weight classification}

We can also use the boost-weight decomposition to classifiy the Weyl tensors 
\cite{class} of 
arbitrary signature. By successively using null-rotations at each level, we can use the well-known boost-weight classification to give an algebraic classification of arbitrary-signature (Weyl) tensors. At each level we can consider the null-rotations leaving invariant the $(2k+m-2I)$-dimensional metric 
\beq
2{\mbold\ell}^I{\mbold n}^I+ \eta_{\mu_I\nu_I}{\mbold\omega}^{\mu_I}{\mbold\omega}^{\nu_I}.
\eeq 
 The metric $\eta_{\mu_I\nu_I}$ will be of signature $(k-I,k-I+m)$. 

Therefore, consider  a Weyl tensor, $C$, which can be decomposed into boost weight components, as explained earlier. To find the primary level algebraic type, we consider the components:
\[ C=(C)_{(+2,\ast,\ast,...,\ast)}+(C)_{(+1,\ast,\ast,...,\ast)}+(C)_{(0,\ast,\ast,...,\ast)}+(C)_{(-1,\ast,\ast,...,\ast)}+(C)_{(-2,\ast,\ast,...,\ast)},\]
where $(+2,\ast,\ast,...,\ast)$ means all the components of 
boost-weight $b_1=+2$, etc. 

We can now use the standard algebraic classification of 
Lorentzian tensors at each level; e.g., will say that $C$ is of 
primary (or primary level) algebraic type III if there is a frame such that $(C)_{(+2,\ast,\ast,...,\ast)}=(C)_{(+1,\ast,\ast,...,\ast)}=(C)_{(0,\ast,\ast,...,\ast)}=0$. 

In order to get the second level type, we use the prefered form from the primary level. Consider the highest non-zero primary boost-weight component $(C)_{(b_1,\ast,\ast,...,\ast)}$. Again, we can decompose as follows:
\[ (C)_{(b_1,\ast,...,\ast)}=\sum_{b_2=-2}^{+2}(C)_{(b_1,b_2,\ast,...,\ast)},\]
The second level type can then be found by trying to find a frame (using the 2th level null-rotations which preserves the primary boost-weights). 

The full algebraic type will then be the sum of the prime,
second, $\dots$. $k$th-level types.   We will write this as follows; e.g., $(I,D,III)$,
means the type at 1st, 2nd, and 3rd level are I, D, and III,
respectively.

For Weyl tensors obeying the ${\bf S}_i$,- or ${\bf N}$-property, we have the following:
\beq
 {\bf S}_i: && (\underbrace{II,II,...,II}_i,G,...,G)\nonumber \\
{\bf N}:  &&({II,II,...,II,III})\nonumber
\eeq
or simpler. 

\section{Four Dimensional Neutral space}

In the special case of  4D neutral signature (NS) space,
the Weyl operator, ${\sf C}$, splits into a self-dual and anti-self-dual part: 
${\sf C}={\sf W}^+\oplus {\sf W}^-$. In \cite{Law} Law classified the Weyl 
tensor of NS metrics using the Weyl operator the (anti-)self-dual operators 
which can be defined as:  
\[ {\sf W}^\pm=\tfrac 12\left({\sf 1}\pm\star\right){\sf C}.\] 
 
Each of the parts can be considered to be symmetric and tracefree with respect to the 3-dimensional Lorentzian metric with signature $(+--)$. Consquently, each of the operators ${\sf W}^{\pm}$ can be classified according to ``Segre type''(the ``Type'' refers to Law's enumeration):  
\begin{itemize} 
\item{} Type Ia: $\{1,11\}$ 
\item{} Type Ib: $\{z\bar{z}1\}$ 
\item{} Type II: $\{21\}$ 
\item{} Type III:$\{3\}$.  
\end{itemize} 
As in \cite{OP} is also advantageous to refine Law's enumeration for the special cases:  
\begin{itemize} 
\item{} Type D: $\{(1,1)1\}$ 
\item{} Type N: $\{(21)\}$  
\end{itemize} 

We first note that the refined Law classification translate into our 
boost-weight classification as follows (for each tensor $W^{\pm}$)
\[ I \leftrightharpoons (I,I), \quad II \leftrightharpoons (II,II)\quad D \leftrightharpoons (D,D)\quad III \leftrightharpoons (III,III)\quad N \leftrightharpoons (N,N).\] 

As for the full Weyl tensor $C=W^++W^-$, there is no simple 1-1 correspondence 
like above. In some sence, there is a ``tilted'' correpondence 
(using a map $G$ in ``boost'' space). In terms of the ${\bf S}^G_i$- and 
${\bf N}$-property, which measures a degeneracy of the Weyl tensor, we can 
relate it to the Law classification as follows.
\begin{prop}
For a 4D neutral signature space ($k=2$, $m=0$), then: 
\begin{enumerate}
\item{} If \emph{either} the self-dual \emph{or} the anti-self-dual part of the Weyl tensor is algebraically special of type II, D, III, N or O, then the Weyl tensor possesses (at least) the ${\bf S}_1^G$-property. 
\item{} If \emph{both}  the self-dual \emph{and} the anti-self-dual part of the Weyl tensor are algebraically special of type II, D, III, N or O, then the Weyl tensor possesses (at least) the ${\bf S}_2^G$-property. 
\item{} If \emph{both}  the self-dual \emph{and} anti-self-dual part of the Weyl tensor are algebraically special of type III, N or O, then the Weyl tensor possesses the ${\bf N}$-property. 
\end{enumerate}
\end{prop}
This result gives us a statement in terms of degeneracy of the Weyl tensor. In particular, it tell us that if one of the parts of the Weyl tensor is algebraically special, then the Weyl tensor is degenerate in the sence that its not determined by its invariants. 
\begin{figure}[tbp]  
\centering \includegraphics[width=10cm]{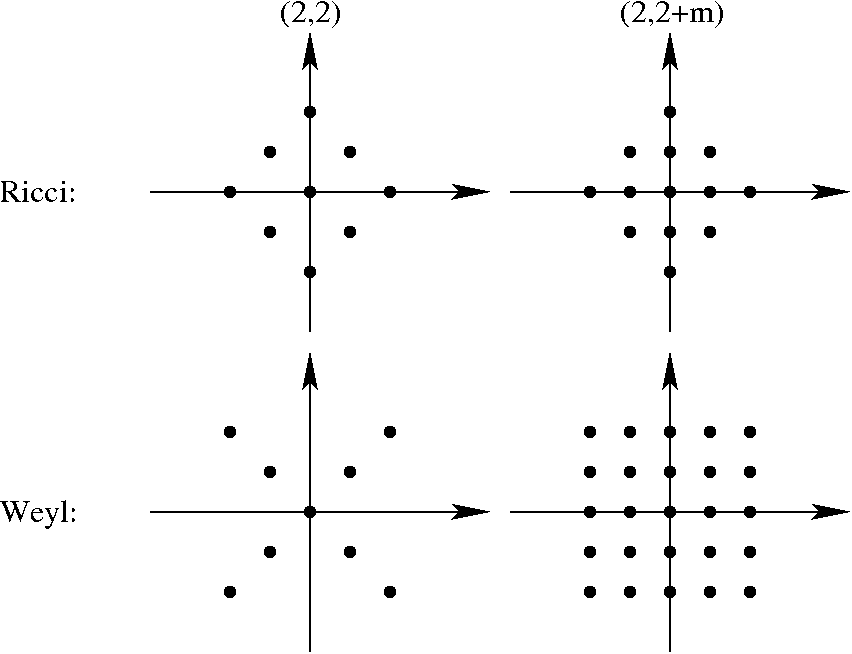}  
\caption{Figures showing the components of the Weyl and Ricci tensor in boost-weight space. Displayed are metrics of signature $(2,2)$ (first column) and $(2,2+m)$, $m>0$ (second column). The axes are the two boost weights $b_1$ and $b_2$. }  
\label{fig1}  
\end{figure} 

In Fig \ref{fig1}, we have illustrated the boost weight components
of the Ricci tensor (the `diamond') and Weyl tensor (the `cross') in 4D
Neutral signature.  Each of the Weyl tensors $W^{\pm}$ are the
diagonal and anti-diagonal components (there are two independent
boost-weight $(0,0)$ components of the Weyl tensor $C$).

Note that for 4D Neutral signature, a Ricci tensor obeying the 
${\bf N}$-property has a 4-step nilpotent Ricci operator: ${\sf R}^4=0$.
\footnote{This need not be the case in higher dimensions. For example, a Ricci tensor ${\sf R}$  of signature 
$(2,2+m)$, $m>0$, possessing the ${\bf N}$-property need not be $4$-step nilpotent, however it must be $5$-step nilpotent; i.e., ${\sf R}^5=0$. The corresponding numbers for the Weyl tensor possessing the ${\bf N}$-property in signatures $(2,2)$ and $(2,2+m)$, $m>0$ are $3$ and $9$ respectively.} By inspection we see that for a Ricci 
tensor obeying the ${\bf N}$-property, we have that the powers of ${\sf R}$ 
must be of the following types (or simplier):
\[ {\sf R}:(II,N), \quad {\sf R}^2:(III,III), \quad{\sf R}^3:(N,D), 
\quad {\sf R}^4:(O,O)\]

As an illustration, let us consider the following  
4D NS spacetime which is a VSI spacetime \cite{VSI}: 
\beq
\d s^2=2\d u^1\left(\d v^1+H\d u^1+W_{\mu_1}\d x^{\mu_1}\right)+2\d u^2\d v^2,
\eeq 
where
\beq
W_{\mu_1}\d x^{\mu_1}&=& v^1W^{(1)}_{u^2}(u^1,u^2)\d u^2+W^{(0)}_{u^2}(u^1,u^2,v^2)\d u^2+W^{(0)}_{v^2}(u^1,u^2,v^2)\d v^2,\nonumber \\
H&=& v^1 H^{(1)}(u^1,u^2,v^2)+H^{(0)}(u^1,u^2,v^2).
\eeq
In general this metric is of Ricci type $(II,N)$.  Here, the term
with $W^{(1)}_{u^2}(u^1,u^2)$ contributes with a boost-weight
$(0,-2)$ component.  Consequently, if $W^{(1)}_{u^2}(u^1,u^2)=0$,
then this spacetime is of type $(III,I)$ and we have ${\sf R}^3=0$.
The above metric also allows for the additional subcases $(III,III)$
and $(N,D)$.  It should be pointed out that the types given here are
not complete in the sense that the metric above allows for the two
distinct cases of $(II,N)$, where ${\sf R}^3\neq 0$ or ${\sf R}^3
=0$.
\section*{Acknowledgments}

The main part of this work was done during a visit to Dalhousie
University April-June 2010 by SH. The work was supported by
NSERC of Canada (AC) and by a Leiv Eirikson mobility grant from the  
Research Council of Norway, project no: {\bf 200910/V11} (SH). 
  
\end{document}